\documentclass{physeauth}
\usepackage{graphicx}
\usepackage{amsmath}
\usepackage{amssymb}
\usepackage{type1cm}

\begin{document}

\begin{frontmatter}

\title{Smeared quantum phase transition in the dissipative random quantum Ising model}

\author[rolla]{Thomas Vojta\thanksref{thank1}},
\author[duke]{Jos\'{e} A. Hoyos}

\address[rolla]{Department of Physics, Missouri University of Science and Technology, Rolla, MO 65409, USA}
\address[duke]{Department of Physics, Duke University, Durham, NC 27708, USA}

\thanks[thank1]{Corresponding author. E-mail: vojtat@mst.edu}

\begin{abstract}
We investigate the quantum phase transition in the random transverse-field Ising model
under the influence of Ohmic dissipation. To this end, we numerically implement a
strong-disorder renormalization-group scheme. We find that Ohmic dissipation destroys
the quantum critical point and the associated quantum Griffiths phase by smearing.
Our results quantitatively confirm a recent theory [Phys. Rev. Lett. {\bf 100},
240601 (2008)] of smeared quantum phase transitions.
\end{abstract}

\begin{keyword}
quantum phase transitions \sep quenched disorder \sep dissipation
\PACS 05.10.Cc \sep 05.70.Fh \sep 75.10.-b
\end{keyword}
\end{frontmatter}

\section{Introduction}

The presence of impurities, defects, or other types of quenched disorder can
qualitatively change the properties of a continuous phase transition. At thermal
transitions, this is controlled by the Harris criterion \cite{Harris74}: If a
clean critical point violates the inequality $\nu > 2/d$ where $\nu$ is the
correlation length exponent and $d$ is the space dimensionality, it is
perturbatively unstable against weak disorder. The generic result of adding
disorder to such a system is a new critical point with different critical
exponents but conventional power-law scaling. Moreover, Griffiths
\cite{Griffiths69} showed that the free energy is singular not just
at the transition, but in an entire parameter region around the transition now
known as the Griffiths phase. This is caused by rare spatial regions that
are locally in the ordered phase while the bulk system is still in the
disordered phase. However, it was soon realized that thermodynamic Griffiths
effects are generically very weak and probably unobservable in
experiment \cite{Imry77}.

At zero-temperature quantum phase transitions, disorder effects can be dramatically
stronger than at thermal transition. Fisher \cite{Fisher92,Fisher95} showed that
the random transverse-field Ising chain has an exotic infinite-randomness critical
point featuring ultraslow activated rather than power-law dynamical scaling. The
associated quantum Griffiths singularities take power-law forms
\cite{ThillHuse95,YoungRieger96}, implying a diverging susceptibility in the Griffiths
phase (for a review on rare region and Griffiths phenomena see, e.g., Ref.\
\cite{Vojta06}).

Dissipation can further enhance the effects of disorder on a quantum phase
transition. For Ising order parameter symmetry, each locally ordered rare region
acts as two-level system which can undergo the localization transition of the
spin-boson problem  when coupled to
an Ohmic bath \cite{LCDFGZ87}. Thus, each region freezes independently of the
bulk system, destroying the sharp phase transition by smearing \cite{Vojta03a}.

Recently, we developed an analytical strong-disorder renormalization-group theory
for the dissipative random transverse-field Ising chain that allowed us to
determine the low-energy fixed points exactly \cite{HoyosVojta08}. We found that
Fisher's infinite randomness critical point \cite{Fisher92,Fisher95} and the
quantum Griffiths phase are indeed destroyed by the dissipation. Instead, there is only
one nontrivial fixed point describing an inhomogeneously ordered phase (the tail of the
smeared transition).

In this paper, we report the results of a numerical implementation of the strong-disorder
renormalization group for the dissipative random transverse-field Ising chain.
The purpose of the work is twofold; (i) it allows us to
confirm and illustrate the theoretical predictions for the asymptotic low-energy
behavior. (ii) More importantly, the numerical simulations allow us to test whether
moderately disordered realistic systems actually flow to the predicted fixed points,
and they allow us to find the relevant crossover scales. The paper is organized as follows.
We introduce the model Hamiltonian in Sec.\ \ref{sec:model}. We summarize the main
results of the analytical renormalization-group theory in Sec.\ \ref{sec:RG}. Section \ref{sec:numerics}
is devoted to the computer simulations. Finally, we conclude in Sec.\
\ref{sec:conclusions}.

\section{Dissipative random transverse-field Ising model}
\label{sec:model}

We consider a one-dimensional random transverse-field Ising model,
a prototypical model displaying an infinite-randomness quantum critical
point. Dissipation is introduced by coupling each spin linearly to an
independent Ohmic bath of harmonic oscillators. The resulting Hamiltonian reads
\begin{eqnarray}
H= & - & \sum_{i}J_{i}\sigma_{i}^{z}\sigma_{i+1}^{z}-\sum_{i}h_{i}\sigma_{i}^{x}\nonumber \\
& + & \sum_{i,n}\sigma_{i}^{z}\lambda_{i,n}(a_{i,n}^{\dagger}+a_{i,n}^{\phantom{\dagger}})+
\sum_{i,n}\nu_{i,n}a_{i,n}^{\dagger}a_{i,n}^{\phantom{\dagger}}~.
\label{eq:H}
\end{eqnarray}
Here, $\sigma_{i}^{x,z}$ are the Pauli matrices representing the spin-1/2 at site $i$.
The bonds $J_{i}$ and transverse fields $h_{i}$ are independent random variables with
bare (initial) probability distributions $P_I(J)$ and $R_I(h)$. $a_{i,n}^{\dagger}$
($a_{i,n}$) are the creation (annihilation) operators of the $n$-th oscillator coupled to
spin $\sigma_{i}$ via $\lambda_{i,n}$, and $\nu_{i,n}$ is its frequency. All baths have
the same bare Ohmic spectral function
\begin{equation}
{\mathcal{E}}(\omega)=\pi\sum_{n}\lambda_{i,n}^{2}\delta(\omega-\nu_{i,n})=2\pi\alpha\omega
e^{-\omega/\omega_{c}}~,
\label{eq:spectral}
\end{equation}
with $\alpha$ the dimensionless dissipation strength and $\omega_{c}$
the (bare) cutoff energy. Under the renormalization group, the cutoff will change,
and the dissipation strength will become site-dependent.

\section{Strong-disorder renormalization group}
\label{sec:RG}

In this section, we briefly summarize our analytical strong-disorder
renormalization-group approach \cite{HoyosVojta08} to the Hamiltonian (\ref{eq:H}). It is
related to a numerical scheme suggested by Schehr and Rieger
\cite{SchehrRieger06,SchehrRieger08}. However, treating the oscillator modes on equal
footing with the spin degrees of freedom allows us to solve the problem analytically.

The basic idea of the strong-disorder renormalization group
\cite{MaDasguptaHu79,DasguptaMa80} is to successively integrate out local high-energy
states. In the Hamiltonian (\ref{eq:H}), the relevant local energies are the transverse
fields $h_i$, the bonds $J_i$, and the oscillator frequencies $\nu_{i,n}$. In more
detail, the renormalization group proceeds as follows. We start by determining the energy
cutoff, i.e., the largest local energy in the system
$\Omega=\max(h_{i},J_{i},\omega_{c}/p)$ where $p\gg1$ is an arbitrary constant. In each
renormalization-group step we now reduce the energy scale from $\Omega$ to
$\Omega-{\textrm{d}}\Omega$ by (i) integrating out all oscillators (at all sites $i$)
with frequencies between $p(\Omega-{\textrm{d}}\Omega)$ and $p\Omega$ using adiabatic
renormalization \cite{LCDFGZ87} and (ii) decimating all transverse fields and all bonds
between $(\Omega-{\textrm{d}}\Omega)$ and $\Omega$ in perturbation theory.

In step (i) all transverse fields renormalize according to
\begin{equation}
\tilde{h}_{i}=h_{i}\left(1-\alpha_{i}\frac{{\textrm{d}}\Omega}{\Omega}\right)
\label{eq:h-tilde-bath}
\end{equation}
while the bonds remain unchanged. Here $\alpha_{i}$ is the renormalized
dissipation strength at site $i$. In step (ii), if two sites are coupled by
a strong bond $J_{i}=\Omega$, the spins $\mathbf{\sigma}_{i}$
and $\mathbf{\sigma}_{i+1}$ can be treated as an effective
spin cluster $\tilde{\mathbf{\sigma}}$ with moment
\begin{equation}
\tilde{\mu}  =  \mu_{i}+\mu_{i+1}~,\label{eq:mu-tilde}
\end{equation}
in a bath of renormalized dissipation strength
\begin{equation}
\tilde{\alpha}=\alpha_{i}+\alpha_{i+1}=\alpha(\mu_{i}+\mu_{i+1})=\alpha\tilde{\mu}~,
\label{eq:alfa-tilde}
\end{equation}
and renormalized transverse field
\begin{equation}
\tilde{h}  =  h_{i}h_{i+1}/J_{i}~.\label{eq:h-tilde-J}
\end{equation}
Conversely, if a site experiences a strong field $h_{i}=\Omega$, the corresponding spin
$\mathbf{\sigma}_{i}$ is eliminated, creating a new bond between sites $i-1$ and $i+1$,
\begin{equation}
\tilde{J}=J_{i-1}J_{i}/h_{i}~.\label{eq:J-tilde}
\end{equation}

We now iterate the complete renormalization-group step (\ref{eq:h-tilde-bath})--(\ref{eq:J-tilde})
decreasing the cutoff energy scale $\Omega$. Using logarithmic
variables $\Gamma=\ln(\Omega_{I}/\Omega)$ {[}where $\Omega_{I}$
is the initial (bare) value of $\Omega${]}, $\zeta=\ln(\Omega/J)$
and $\beta=\ln(\Omega/h)$, we can derive renormalization-group flow equations for
the probability distribution ${\mathcal{P}}(\zeta)$ of the bonds and the joint distribution
${\mathcal{R}}(\beta,\mu)$ of the fields and moments.
They are given by
\begin{eqnarray}
\frac{\partial{\mathcal P}}{\partial\Gamma} & = & \frac{\partial{\mathcal P}}{\partial\zeta}+\left(1-\alpha\overline{\mu}_{0}\right){\mathcal R}_{\beta}\left(0\right)\left({\mathcal P}\stackrel{\zeta}{\otimes}{\mathcal P}\right)\nonumber \\
 &  & +\left[{\mathcal P}\left(0\right)-\left(1-\alpha\overline{\mu}_{0}\right){\mathcal R}_{\beta}\left(0\right)\right]{\mathcal P}~,\label{eq:flow-P}\\
\frac{\partial{\mathcal R}}{\partial\Gamma} & = & \left(1-\alpha\mu\right)\frac{\partial{\mathcal R}}{\partial\beta}+{\mathcal P}\left(0\right)\left({\mathcal R}\stackrel{\beta,\mu}{\otimes}{\mathcal R}\right)\nonumber \\
 &  & -\left[{\mathcal P}\left(0\right)-\left(1-\alpha\overline{\mu}_{0}\right){\mathcal R}_{\beta}\left(0\right)\right]{\mathcal R}~,\label{eq:flow-R}
\end{eqnarray}
where ${\mathcal{R}}_{\beta}(\beta)=\int_{0}^{\infty}{\mathcal{R}}(\beta,\mu){\textrm{d}}\mu$
is the distribution of the fields and $\overline{\mu}_{0}$ is the
average moment of clusters about to be decimated (defined by $\overline{\mu}_{0}{\mathcal{R}}_{\beta}(0)=\int_{0}^{\infty}\mu{\mathcal{R}}(0,\mu){\textrm{d}}\mu$).
The symbol ${\mathcal{P}}\stackrel{\zeta}{\otimes}{\mathcal{P}}=\int_{0}^{\zeta}{\mathcal{P}}(\zeta^{\prime}){\mathcal{P}}(\zeta-\zeta^{\prime}){\textrm{d}}\zeta^{\prime}$
denotes the convolution. The first term on the r.h.s. of (\ref{eq:flow-P})
and (\ref{eq:flow-R}) is due to the rescaling of $\zeta$ and $\beta$
with $\Gamma$ and the renormalization (\ref{eq:h-tilde-bath}) of
$h$ by the baths. The second term implements the recursion relations
(\ref{eq:mu-tilde}), (\ref{eq:h-tilde-J}) and (\ref{eq:J-tilde}) for the
moments, fields and bonds. The last term ensures the normalization
of $\mathcal{P}$ and $\mathcal{R}$. As expected, for $\alpha=0$,
(\ref{eq:flow-P}) and (\ref{eq:flow-R}) become identical to the
dissipationless case \cite{Fisher92,Fisher95}.

The qualitative change of the physics brought about by the dissipation can be seen already
in the probability of decimating a field,
$(1-\alpha\overline{\mu}_{0}){\mathcal{R}}_{\beta}(0)$,
which decreases with increasing dissipation strength and cluster size. Clusters
with moment $\mu>1/\alpha$ are never decimated. Thus, in the presence
of dissipation, the flow equations contain a finite length scale above which
the cluster dynamics freezes.

In Ref.\ \cite{HoyosVojta08}, we determined the fixed points, i.e., stationary solutions
of the flow equations (\ref{eq:flow-P}) and (\ref{eq:flow-R})
(invariant under a general rescaling $\eta=\zeta/f_{\zeta}(\Gamma)$,
$\theta=\beta/f_{\beta}(\Gamma)$ and $\nu=\mu/f_{\mu}(\Gamma)$). They correspond to
stable phases or critical points. We found that the dissipation destroys the
the critical fixed point of the dissipationless model
\cite{Fisher92,Fisher95} and the ones associated with the corresponding
disordered and ordered quantum Griffiths phases. Instead, for overlapping bond and field distributions,
there is only one line of well-behaved fixed points (parameterized by
${\mathcal{P}}_{0}>0$) corresponding to the tail of the ordered phase.
Here, $f_{\zeta}=1$, $f_{\mu}=\exp({\mathcal{P}}_{0}\Gamma)$,
$f_{\beta}=\Gamma\exp({\mathcal{P}}_{0}\Gamma)$. The fixed-point
distributions can be found in closed form, they read
\begin{equation}
{\mathcal{P}}^{*}(\zeta) = {\mathcal{P}}_{0}e^{-{\mathcal{P}}_{0}\zeta}~,\label{eq:fixed-point-p}
\end{equation}
\begin{equation}
{\mathcal{R}}^{*}(\theta,\nu) = {\mathcal{R}}_{0} 
e^{-{\mathcal{R}}_{0}\nu} \delta(\theta - \alpha \nu )~,\label{eq:fixed-point-r}
\end{equation}
with ${\mathcal{R}}_{0}$ being a nonuniversal constant. This fixed point is similar to the
ordered Griffiths phase in the dissipationless case, but $f_{\beta}/f_{\mu}\rightarrow\infty$
as $\Gamma\rightarrow\infty$. Transforming the field distribution
(\ref{eq:fixed-point-r}) back to the original transverse fields $h$
gives power-law behavior $\sim h^{{\mathcal{R}}_{0}/(\alpha f_{\beta})-1}$.
This result implies that there is no fixed-point solution
with $f_{\zeta}/f_{\beta}\rightarrow{\textrm{const}}$ as $\Gamma\rightarrow\infty$
in the presence of dissipation, proving that there is no quantum critical point
where fields and bonds compete at \emph{all} energy scales.

\section{Numerical results}
\label{sec:numerics}

In this main section of the paper, we present numerical renormalization-group results for
the Hamiltonian (\ref{eq:H}) with moderate disorder. The goal is to check whether
the renormalization-group flow is indeed towards the fixed point
(\ref{eq:fixed-point-p},\ref{eq:fixed-point-r}) and to investigate the crossover
from dissipationless to dissipative behavior for small dissipation strength.

To do so, we numerically implement a strong-disorder renormalization-group
scheme similar to Refs.\ \cite{SchehrRieger06,SchehrRieger08}, and apply it to very
long chains up to $L_0=10^6$ sites. All data are averages over $10^3$ different disorder
realizations. The random bonds and fields are drawn from probability distributions
\begin{equation}
P_I(J) \propto J^{-x} \qquad (J_{\rm min}<J<J_{\rm max})~,
\label{eq:PI-J}
\end{equation}
\begin{equation}
R_I(h) \propto h^{-x} \qquad (h_{\rm min}<h<h_{\rm max})~.
\label{eq:RI-h}
\end{equation}
All sites have the same initial magnetic moment $\mu_i = 1$ and oscillator cutoff
$\omega_c = p h_{\rm max}$. We have used $p=1$, but because $p$ does not appear in
the recursion relations (\ref{eq:h-tilde-bath})--(\ref{eq:J-tilde}), using any other
value just amounts to a redefinition of the bare distribution $R_I(h)$. Using this
method, we have investigated a large number of different parameter sets. In following we
present a selection of typical results.

\begin{figure}[tb]
  \begin{center}
    \includegraphics[angle=0,width=.4\textwidth]{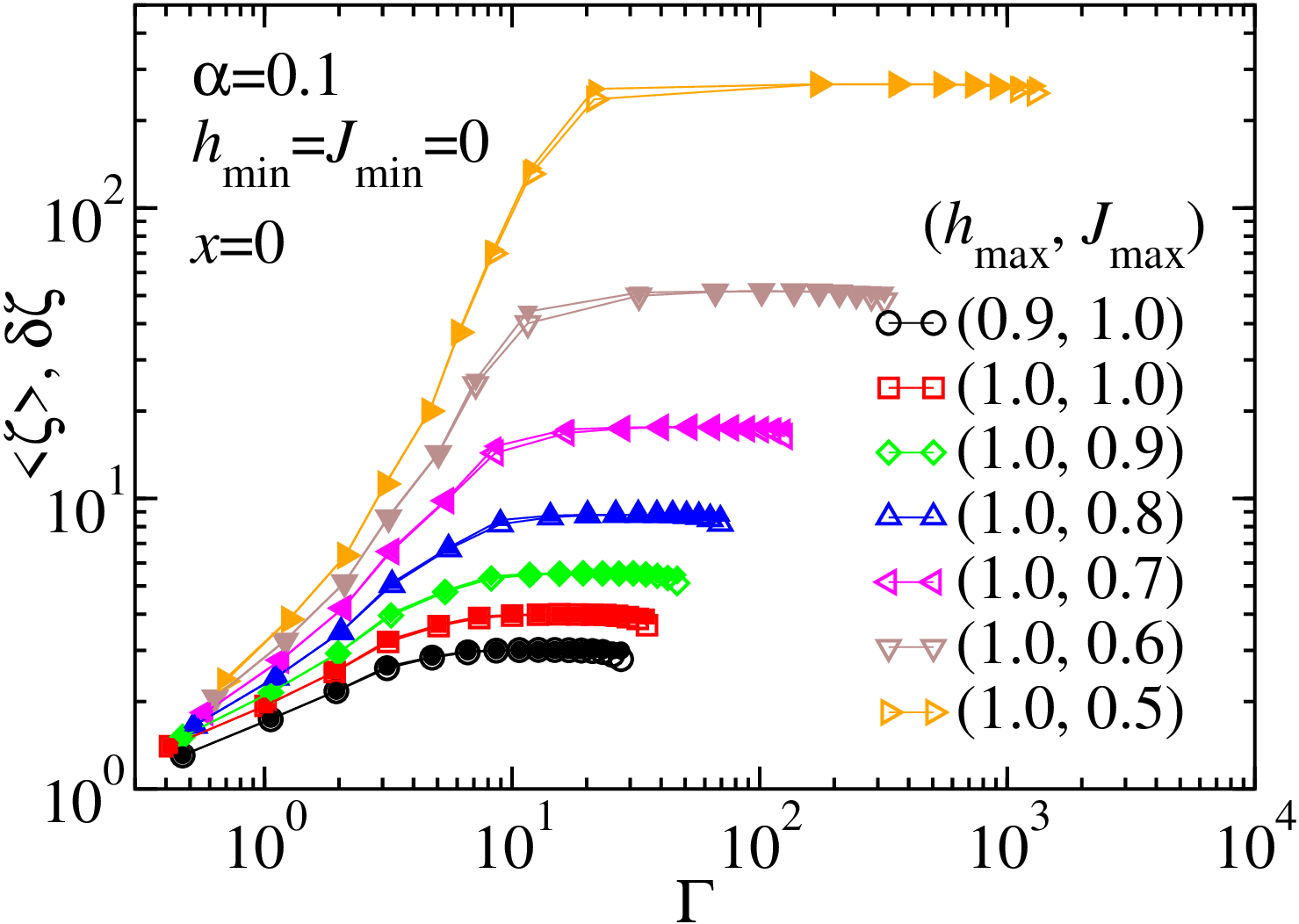}
    \caption{Renormalization-group flow of $\langle \zeta \rangle$ (open symbols) and $\delta\zeta$ (filled symbols) as a function
    of the cutoff energy scale, $\Gamma=\ln(\Omega_I/\Omega)$ for several different chains.
    Note that $\langle \zeta \rangle$ and  $\delta\zeta$ agree to good approximation.}
    \label{fig:zeta}
  \end{center}
\end{figure}

Let us start by considering the renormalization-group flow of the averages and standard deviations
of the logarithmic bond and field variables as well as the magnetic moment. According to
the fixed-point solution (\ref{eq:fixed-point-p},\ref{eq:fixed-point-r}), they are
expected to behave as
\begin{equation}
\langle\zeta\rangle=\delta\zeta=1/{\mathcal P}_0~,
\label{eq:zeta_av}
\end{equation}
\begin{equation}
\langle\beta\rangle=\delta\beta=\alpha\Gamma e^{\mathcal P_0 \Gamma}/{\mathcal R}_0~,
\label{eq:beta_av}
\end{equation}
\begin{equation}
\langle\mu\rangle=\delta\mu=e^{\mathcal P_0 \Gamma}/{\mathcal R}_0
\label{eq:mu_av}
\end{equation}
in the low-energy limit $\Gamma \to \infty$.
In Fig.\ \ref{fig:zeta}, we show the evolution of the average and standard deviation of the logarithmic bond
variable $\zeta$ for several chains with different $h_{\rm max}$ and $J_{\rm max}$.\footnote{Before
measuring any quantity we always integrate out the oscillators such that
all local cutoffs $\omega_{c,i}$ are identical.}
The dissipation strength is fixed at $\alpha=0.1$.
Without dissipation, the chain with $(h_{\rm max},J_{\rm max})=(0.9,1.0)$ would be in the
ordered phase, the chain with (1.0,1.0) would be critical, and all others would be in the
disordered phase. The figure shows that $\langle \zeta \rangle$ and $\delta\zeta$
initially increase under renormalization (corresponding to a rapid drop of the bond energies $J$);
however after a sharp crossover, they settle on a constant value as expected from (\ref{eq:zeta_av}).

\begin{figure}[tb]
  \begin{center}
    \includegraphics[angle=0,width=.4\textwidth]{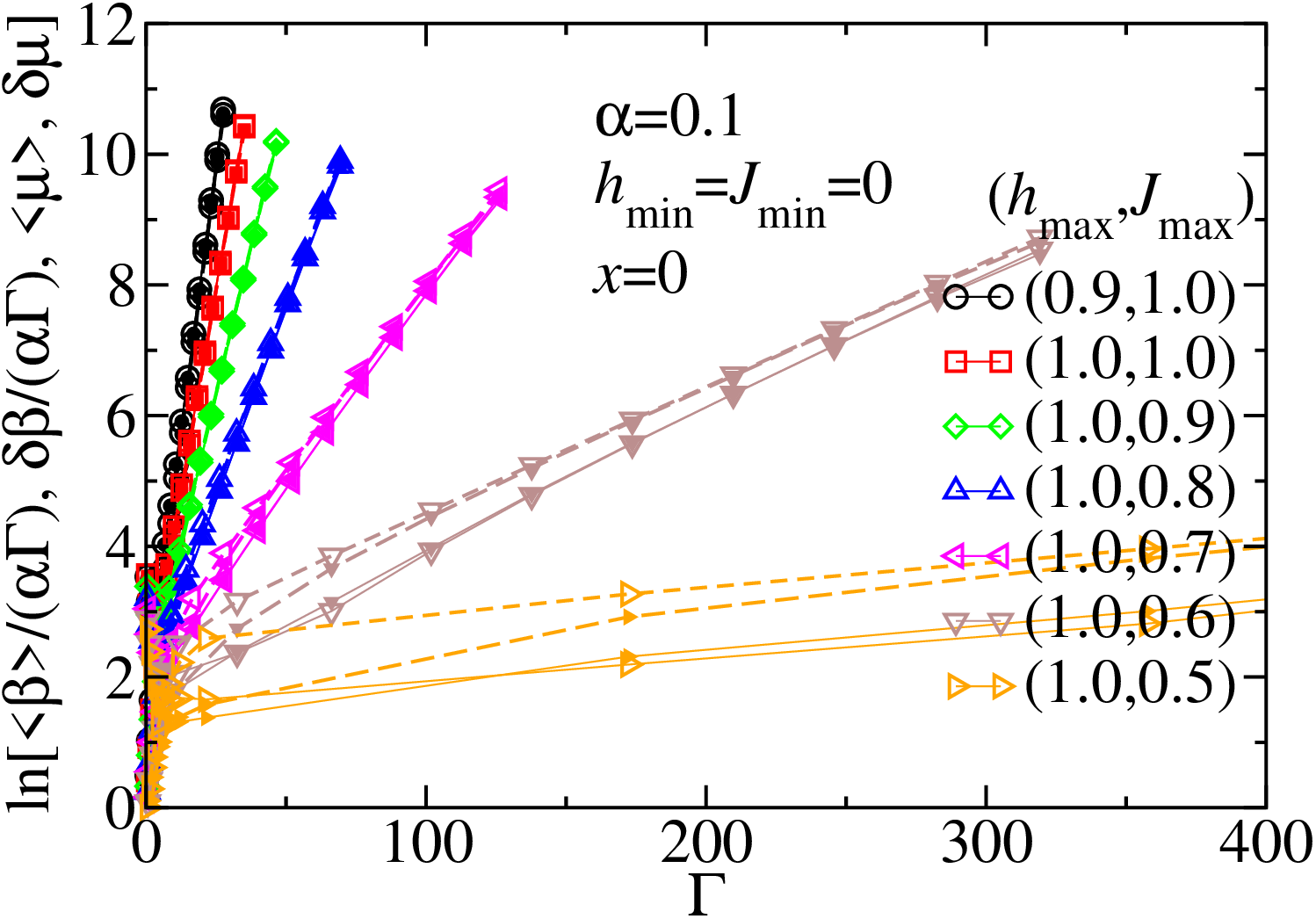}
    \caption{Renormalization-group flow of
    $\langle \beta \rangle$, $\langle \mu \rangle$ (open symbols) and $\delta\beta$, $\delta\mu$ (filled symbols) as a function
    of the cutoff energy scale, $\Gamma=\ln(\Omega_I/\Omega)$. The solid lines represent $\langle \beta \rangle$ and $\delta\beta$ while the
    dashed lines represent $\langle \mu \rangle$ and $\delta\mu$. Again, averages and standard deviations agree very well
    for $\Gamma \to \infty$.}
    \label{fig:beta-mu}
  \end{center}
\end{figure}

Figure \ref{fig:beta-mu} shows the corresponding data for the averages of the logarithmic field variable
$\beta$, the cluster magnetic moment $\mu$, and their respective standard deviations.
They are plotted as $\ln(\langle\beta\rangle/(\alpha\Gamma))$, $\ln(\langle\mu\rangle)$,
$\ln(\delta\beta/(\alpha\Gamma))$, and $\ln(\delta\mu)$ versus
$\Gamma$. In this plot, the fixed-point forms (\ref{eq:beta_av}), (\ref{eq:mu_av})
correspond to straight lines. The figure shows that all data indeed follow the expected
behavior after some short initial transients.

\begin{figure}[tb]
  \begin{center}
    \includegraphics[angle=0,width=.4\textwidth,clip]{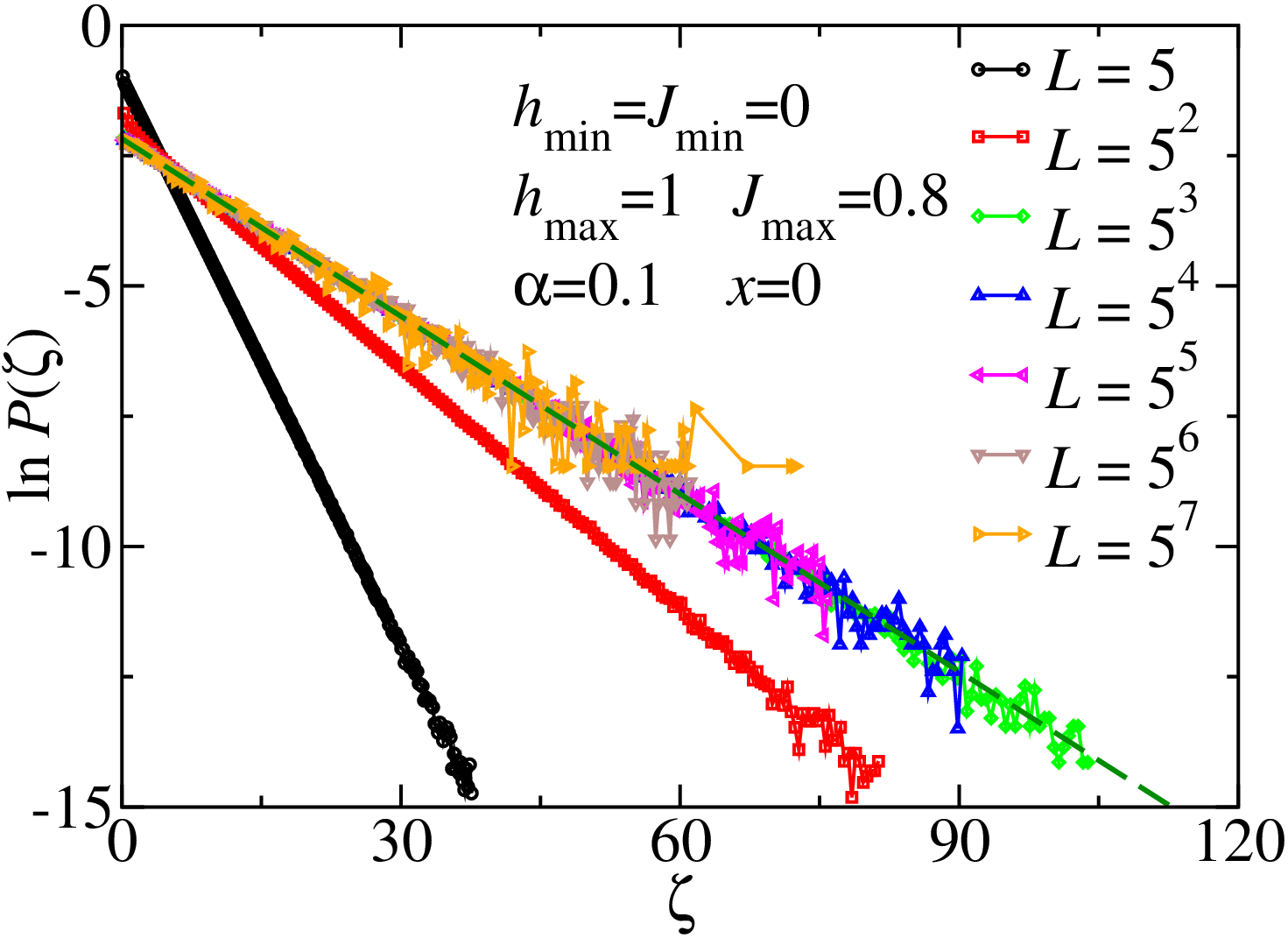}
    \includegraphics[angle=0,width=.4\textwidth,clip]{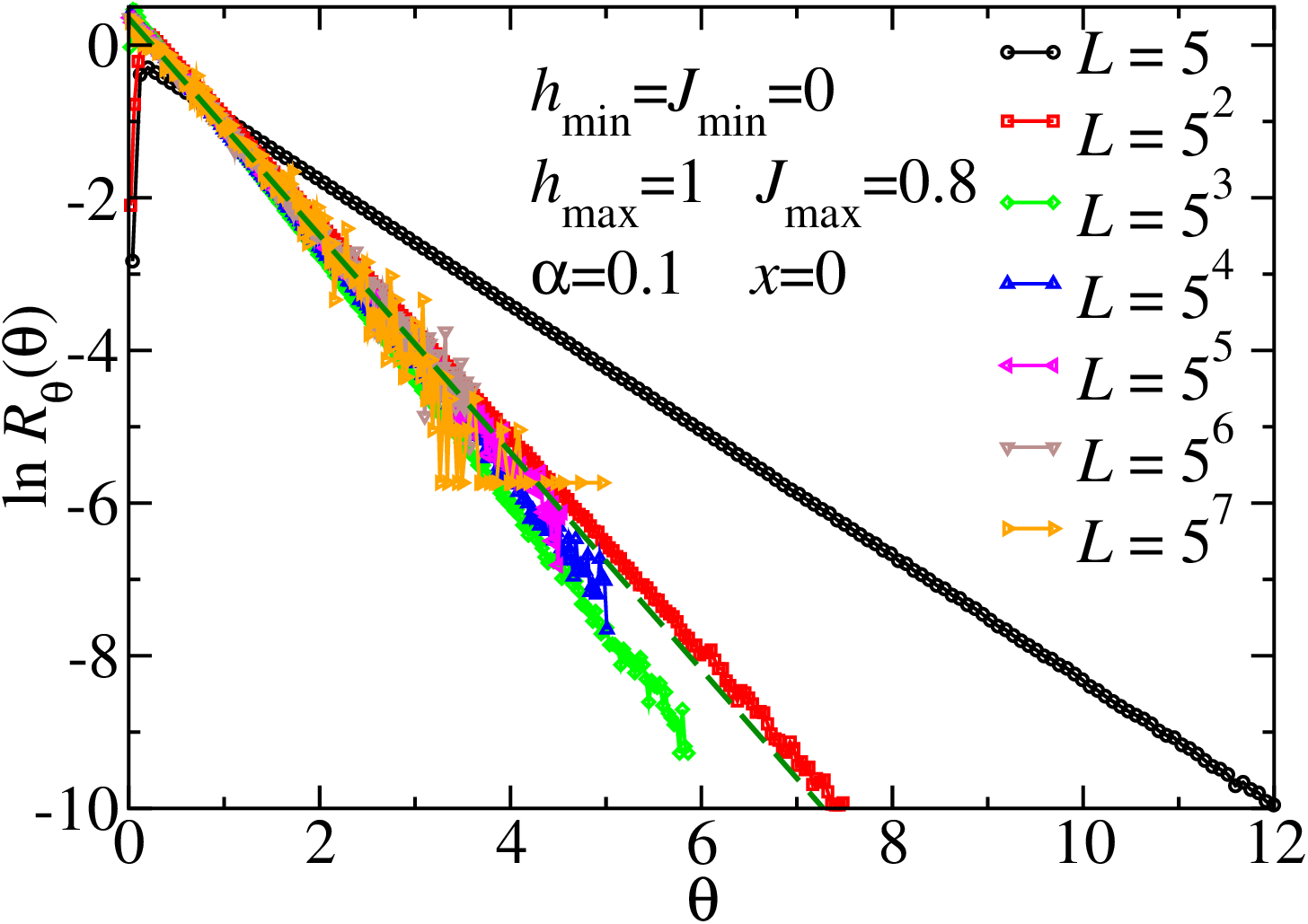}
    \includegraphics[angle=0,width=.4\textwidth,clip]{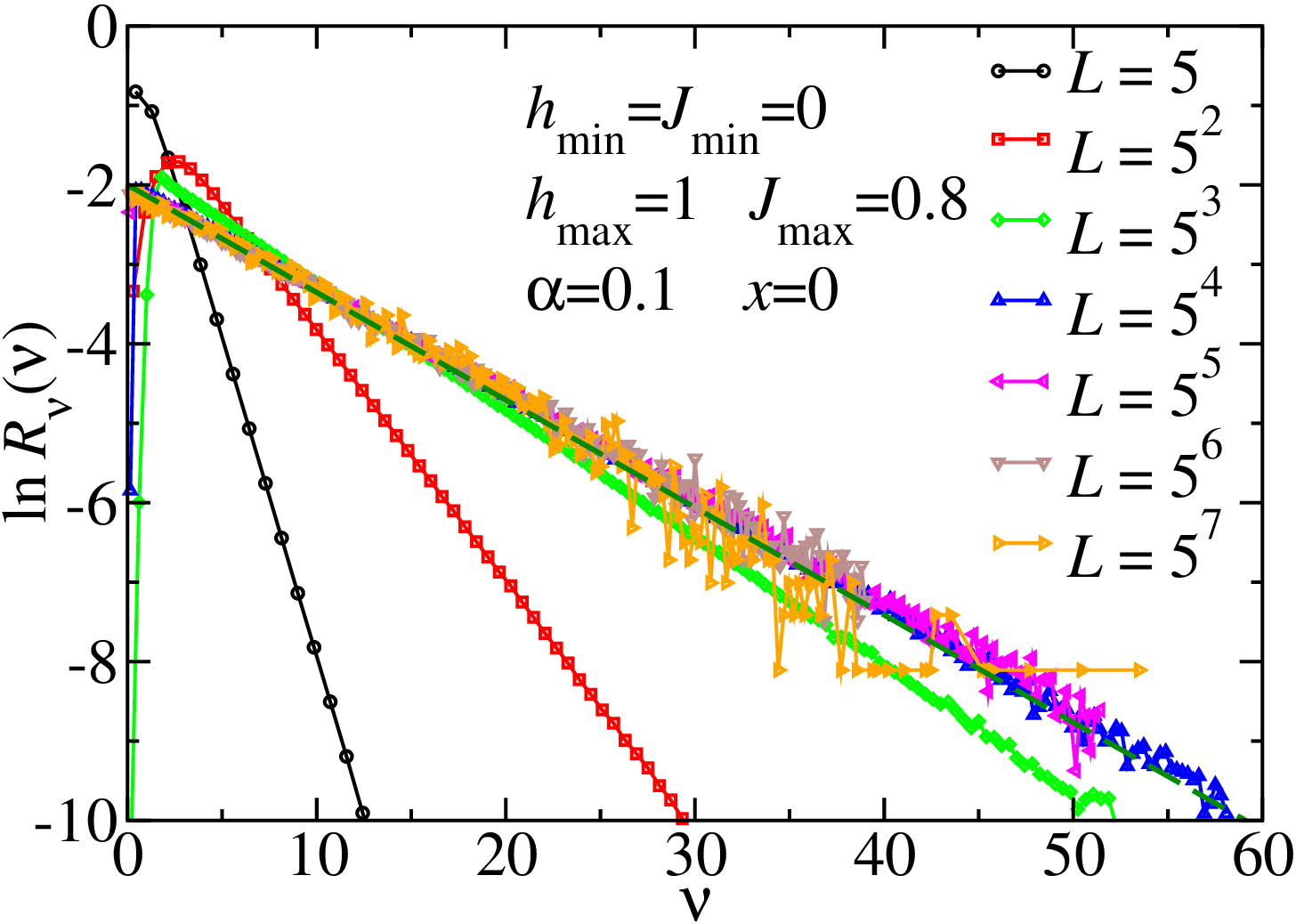}
    \caption{Probability distributions $\mathcal{P}(\zeta)$, $\mathcal{R}_\theta(\theta)$ and
        $\mathcal{R_\nu}(\nu)$ taken at different stages of the renormalization-group flow
        parametrized by the average distance $L$ of the remaining spin clusters. The dashed
        lines represent the fixed-point distributions (\ref{eq:fixed-point-p}) and
        (\ref{eq:fixed-point-r}). $(h_{\rm max},J_{\rm max})=(1.0,0.8)$, the parameters are
        as in Fig.\ \ref{fig:zeta} and \ref{fig:beta-mu}.}
    \label{fig:distribs}
  \end{center}
\end{figure}

So far, we have analyzed the renormalization-group flow of the averages and standard deviations
of the bond and field variables as well as the magnetic moments. For a full confirmation
of the theoretical predictions (\ref{eq:fixed-point-p}) and (\ref{eq:fixed-point-r}), we
need to analyze the full probability distributions. Figure \ref{fig:distribs} shows
snapshots of the probability distributions $\mathcal{P}(\zeta)$, $\mathcal{R}_\theta(\theta)$ and
$\mathcal{R_\nu}(\nu)$ along the renormalization-group flow.
Here, we have used
the rescaled field variable $\theta=\beta/(\Gamma e^{\mathcal{P}_0 \Gamma})$  and the
rescaled moment $\nu=\mu/(\Gamma e^{\mathcal{P}_0})$. The figure shows that
all distributions quickly approach the predicted functional forms. In the late
renormalization-group stages, the numerical data are in excellent quantitative agreement
with the theoretical results (\ref{eq:fixed-point-p}) and (\ref{eq:fixed-point-r}).

\begin{figure}[tb]
  \begin{center}
    \includegraphics[angle=0,width=.4\textwidth,clip]{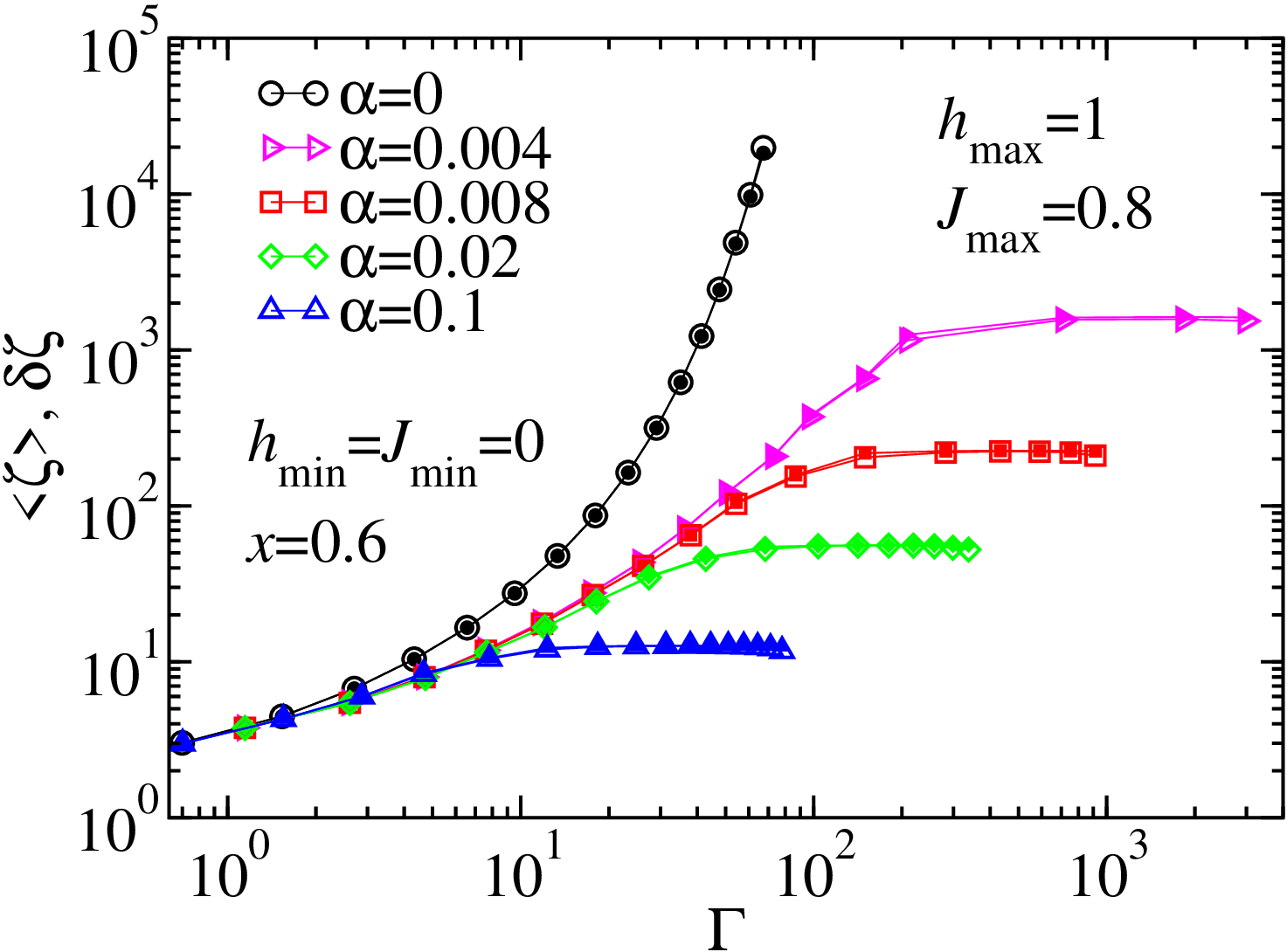}
    \includegraphics[angle=0,width=.4\textwidth,clip]{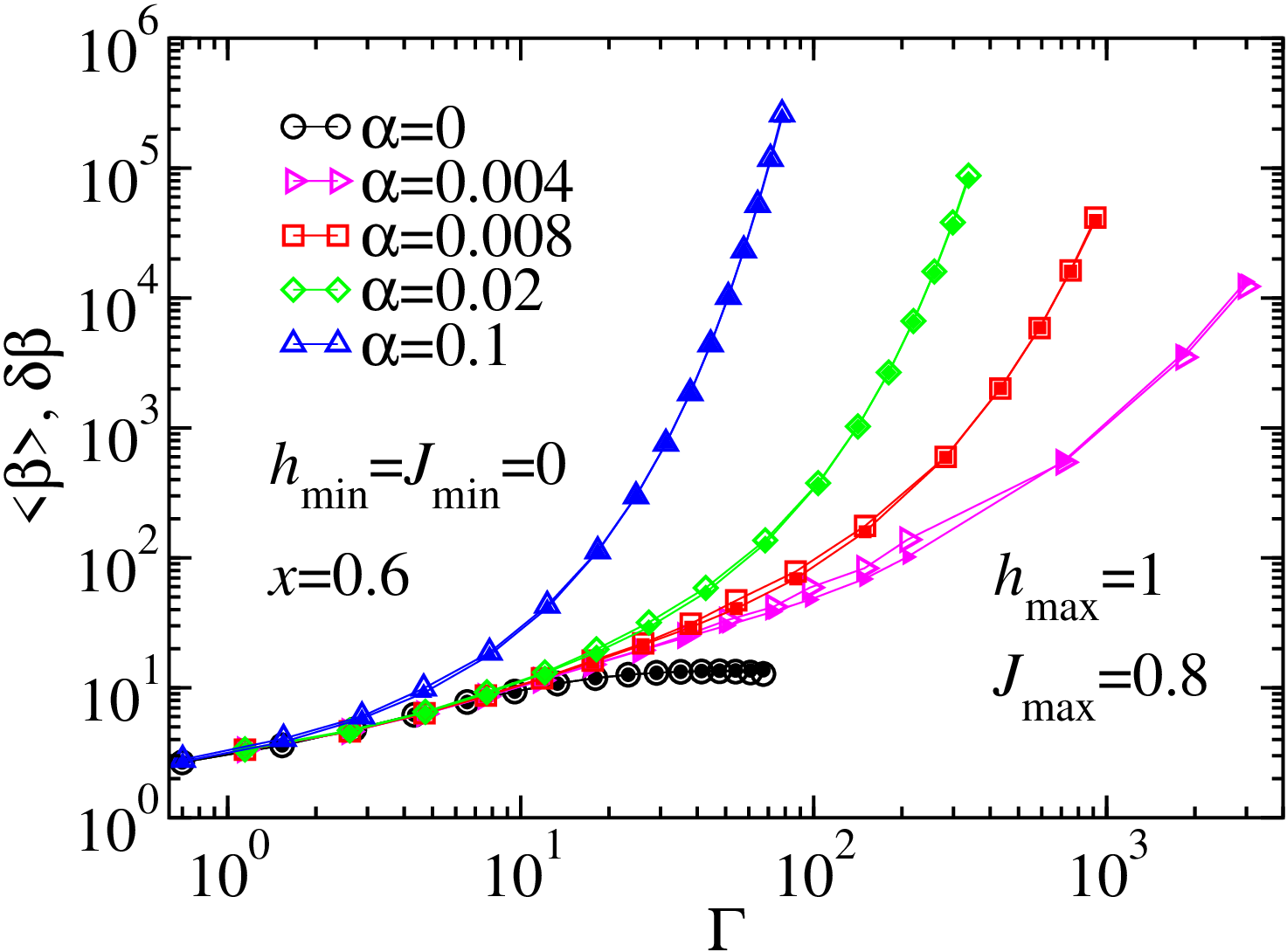}
    \includegraphics[angle=0,width=.4\textwidth,clip]{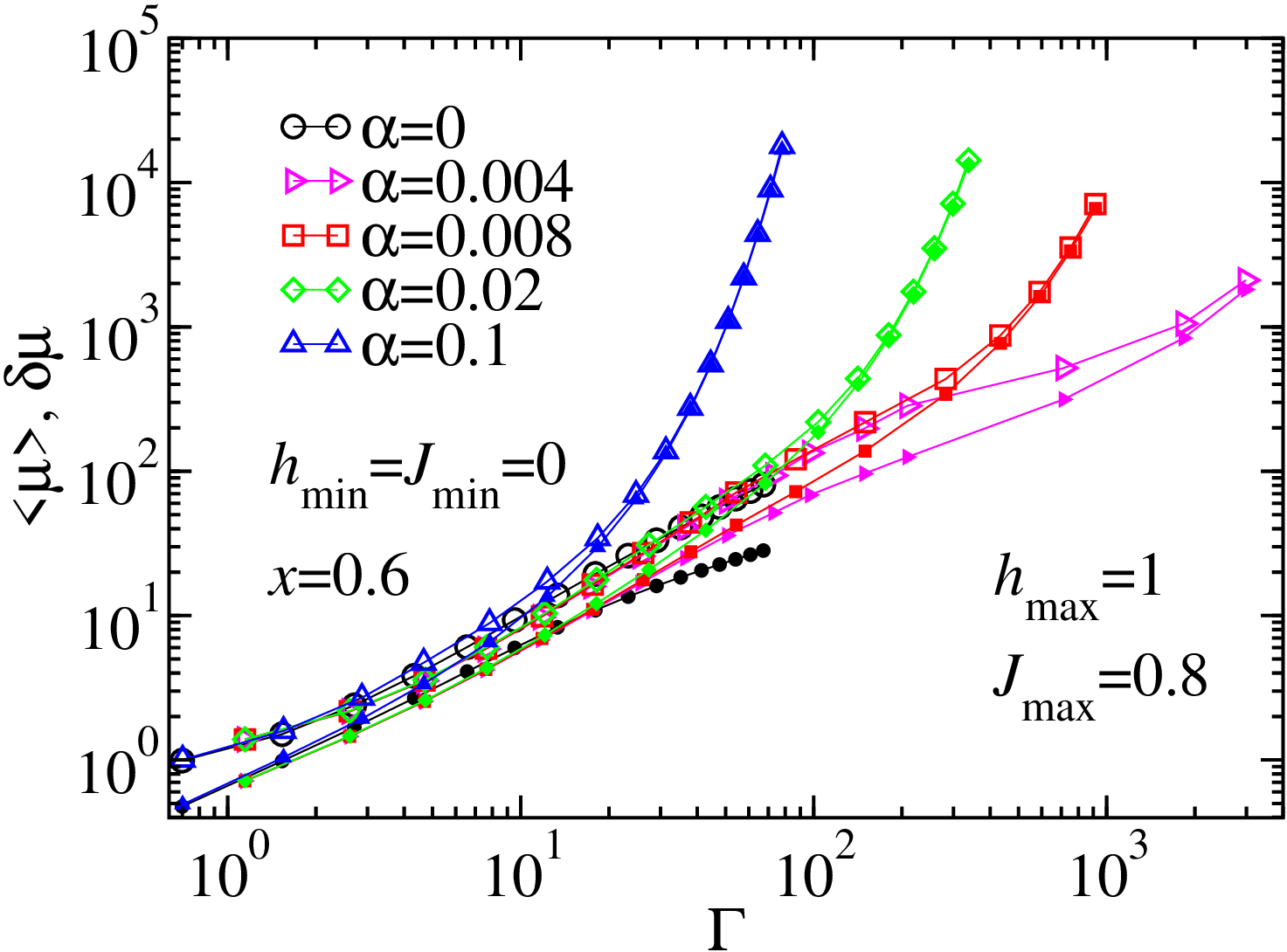}
    \caption{Renormalization-group flow of $\langle \zeta \rangle$,
    $\langle \beta \rangle$, and $\langle \mu \rangle$ (open symbols) as well as their standard
    deviations (filled symbols) as a function of the cutoff energy scale,
    $\Gamma=\ln(\Omega_I/\Omega)$ for $(h_{\rm max},J_{\rm max})=(1.0,0.8)$.}
    \label{fig:alpha_flow}
  \end{center}
\end{figure}

After having analyzed the behavior at fixed value of the dissipation strength $\alpha$,
we now turn to the $\alpha$-dependence of the renormalization-group flow. Figure
\ref{fig:alpha_flow} shows the evolution of  $\langle \zeta \rangle$,  $\langle \beta
\rangle$, and  $\langle \mu \rangle$ as well as their standard deviations for different
$\alpha$.
Because $h_{\rm max}=1.0$ is larger than $J_{\rm max}=0.8$, the dissipationless
chain ($\alpha=0$) is in the disordered (Griffiths) phase. This can be seen from the
rapid increase of the bond variable $\langle \zeta \rangle$, corresponding to a rapid drop of the
interaction energies $J$ under renormalization while the field variable $\langle \beta \rangle$
approaches a constant and the moment $\langle \mu \rangle$ of the surviving
clusters increases linearly with $\Gamma$ \cite{Fisher92,Fisher95}.

In the presence of
dissipation, the flow changes qualitatively when the typical cluster size reaches
$1/\alpha$, as can be seen in the third panel of Fig.\ \ref{fig:alpha_flow}. Beyond this
crossover scale, it is now the bond variable $\langle \zeta \rangle$ that saturates
while the field variable $\langle \beta \rangle$ and the moment $\langle \mu \rangle$
rapidly increase, in agreement with Eqs.\
(\ref{eq:zeta_av}), (\ref{eq:beta_av}), and (\ref{eq:mu_av}).
Thus, at a cluster moment of $1/\alpha$ the
flow character changes from that of the disordered Griffiths phase to that of our
inhomogeneously ordered phase. This crossover is caused by the fact that clusters
with $\alpha\mu>1$ undergo the localization transition of the
spin-boson problem \cite{LCDFGZ87} and are not decimated under further renormalization.

\begin{figure}[tb]
  \begin{center}
    \includegraphics[angle=0,width=.4\textwidth,clip]{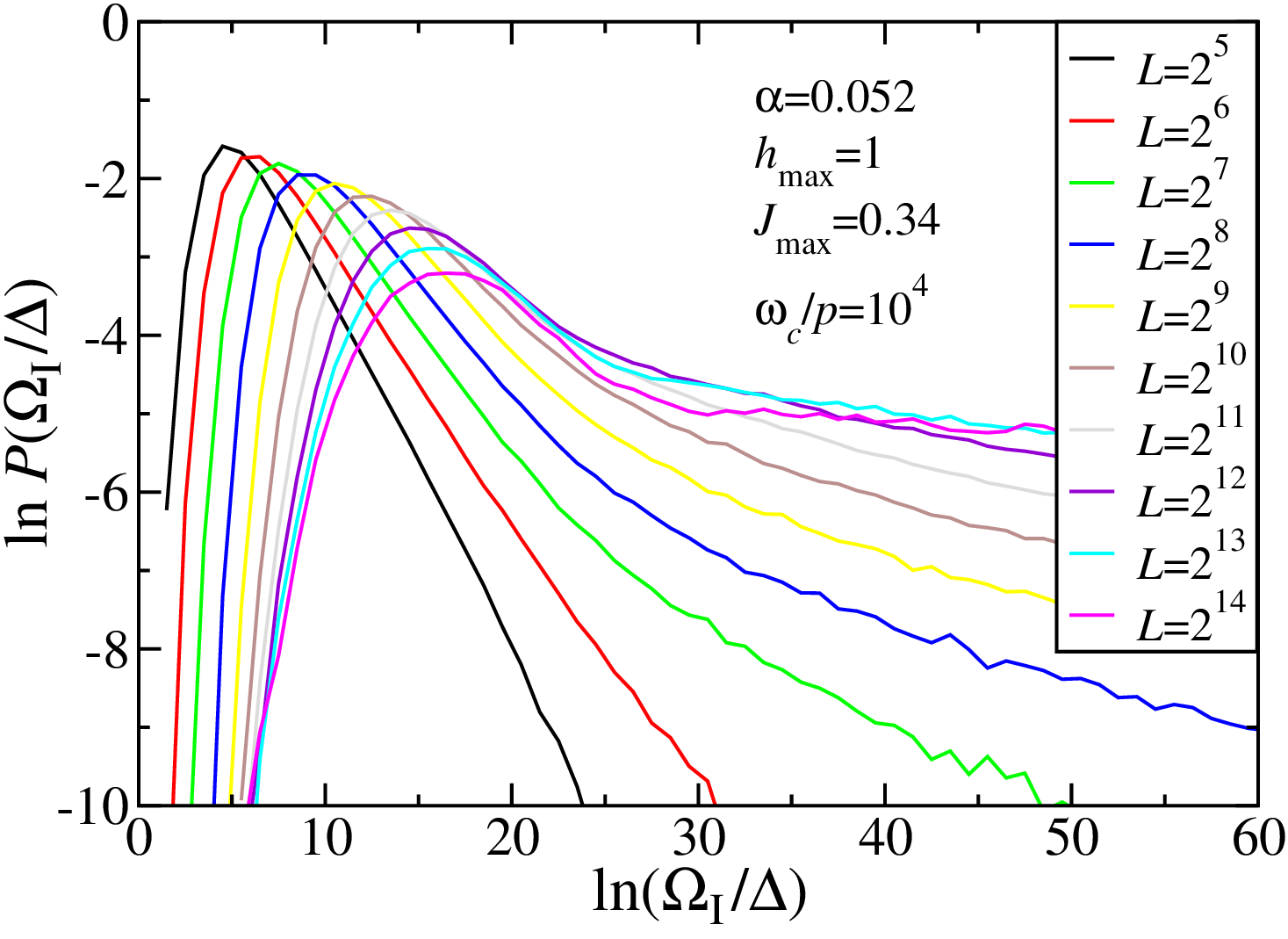}
    \includegraphics[angle=0,width=.4\textwidth,clip]{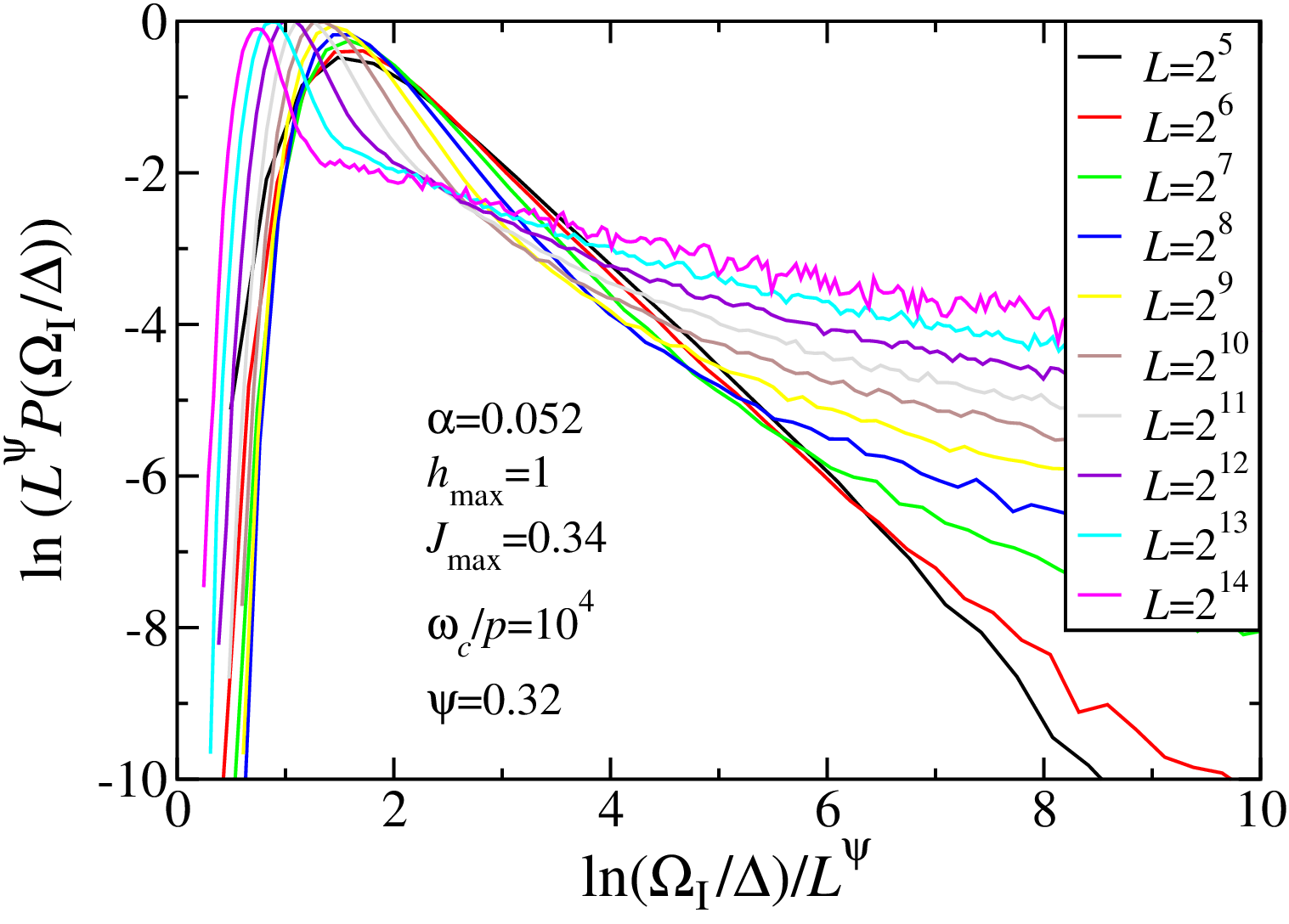}
    \caption{Probability distribution of the smallest energy gap $\Delta$ of a finite-size chain
    for different chain lengths. The parameters $h_{\rm max}=1.0$, $J_{\rm max}=0.34$,
    $h_{\rm min}=J_{\rm min}=0$,
    $x=0$, and $\alpha=0.052$ correspond to the pseudo-critical point of Ref.\ \cite{SchehrRieger06}.}
    \label{fig:pseudo}
  \end{center}
\end{figure}

Finally, we have also studied the ultimate fate of the pseudo-critical point identified
at intermediate scales by Schehr and Rieger. To this end, we have
repeated the calculation of Ref.\ \cite{SchehrRieger06} for much longer chains of up to
16000 sites (averaging over 10$^6$ disorder realizations).
The quantity studied is the smallest excitation energy $\Delta$ of a
finite chain, as estimated by the last nonzero transverse-field in the renormalization
procedure. The upper panel of Fig.\ \ref{fig:pseudo} shows the probability distribution of
$\ln(\Omega_I/\Delta)$ for different chain sizes.
Following Ref.\ \cite{SchehrRieger06}, only
chains that are not yet frozen ($\Delta \ne 0$) at the end of the renormalization procedure
are included in the distribution. The parameters
$h_{\rm max}=1.0$, $J_{\rm max}=0.34$, $x=0$, and $\alpha=0.052$ exactly correspond to the
pseudo-critical point of \cite{SchehrRieger06}.
The figure shows that the character of the distribution changes significantly with
increasing chain length $L$. The lower panel demonstrates that the data for $L\lessapprox 512$
and not too large $\ln(\Omega_I/\Delta)$ can be approximately scaled according to the
pseudo-critical activated scaling form $\ln(\Omega_I/\Delta) \sim L^\psi$ with
$\psi=0.32$. However, for longer chains the distributions do not scale. Instead, they
become much broader, indicating a crossover to the functional form
(\ref{eq:fixed-point-r}) characterizing the inhomogeneously ordered phase. Thus, the
pseudo-critical behavior applies only to a transient regime of the renormalization-group
flow and thus only to a transient energy or temperature window \cite{SchehrRieger08}.

\section{Conclusions}
\label{sec:conclusions}

In summary, we have presented results of an extensive numerical strong-disorder
renormalization group for the random transverse-field Ising model with Ohmic
dissipation. Our simulations quantitatively confirm the analytical theory of Ref.\
\cite{HoyosVojta08}. Specifically, we have verified that the Ohmic dissipation
destabilizes the quantum critical of the dissipationless system
and the associated quantum Griffiths phase. No other critical point has been
found. Instead, the low-energy behavior
in the region of overlapping field and bond distributions is governed by
a new line of fixed points describing the inhomogeneous tail of the ordered
phase. Thus, the sharp quantum phase transition is destroyed by smearing
due to the \emph{interplay} of disorder and Ohmic dissipation.
Note that Ohmic dissipation also suppresses the quantum Griffiths singularities
at the percolation quantum phase transition \cite{SenthilSachdev96}
in a \emph{diluted} transverse-field Ising model \cite{HoyosVojta06}.
However, the percolation transition remains sharp because it is driven
by the critical geometry of the lattice.

In addition to confirming the fixed-point structure of the analytical theory
\cite{HoyosVojta08},  our extensive numerical results also show that moderately
disordered systems generically flow towards the new line of fixed points. The crossover
to the asymptotic behavior occurs when the typical cluster moment reaches $1/\alpha$.

Let us conclude by putting our results into a broader perspective.
Recently, a general classification has been put forward of phase transitions
in the presence of weak disorder \cite{Vojta06}. It is based on the effective
dimensionality of the defects or rare regions. Three classes need to be distinguished.
(i) If the defect dimension is below the lower critical dimension $d_{c}^{-}$
of the problem, the behavior is conventional; (ii) if it is right at $d_{c}^{-}$,
the transition is of infinite-randomness type; and (iii) if it is above
$d_{c}^{-}$, finite clusters can order independently leading to a
smeared transition. In our case, individual rare regions can undergo the
localization transition of the spin-boson problem \cite{LCDFGZ87}.
The system therefore falls into the smeared-transition class.

The results for a dissipative \emph{Ising} magnet must be contrasted
with the behavior of systems with \emph{continuous} O($N$) symmetry.
While large Ising clusters freeze in the presence of Ohmic dissipation,
O($N$) clusters continue to fluctuate with a rate exponentially small
in their moment \cite{VojtaSchmalian05}, putting the system into class
(ii). This leads to a sharp
transition controlled by an infinite-randomness critical point in
the same universality class as the dissipationless random transverse-field
Ising model \cite{HoyosKotabageVojta07,VojtaKotabageHoyos08}.

Our results directly apply to quantum phase transitions in disordered systems
with discrete order parameter symmetry and Ohmic dissipation. The renormalization-group
approach should be broadly applicable to a variety of disordered dissipative
quantum systems such as arrays of resistively shunted Josephson junctions.


\section*{Acknowledgments}

We acknowledge stimulating discussions with H. Rieger, G. Schehr, and F. Igloi.
This work was supported by the NSF under Grants Nos. DMR-0339147 and
DMR-0506953, by Research Corporation, and by the University of Missouri
Research Board.

%

\end{document}